\documentclass[usegraphicx]{mn2e}
\def\ms{M_{\odot}}
\begin{document}
\title[MOND Galaxy formation]{Forming galaxies with MOND}
\author[R.H. Sanders] {R.H.~Sanders\\Kapteyn Astronomical Institute,
P.O.~Box 800,  9700 AV Groningen, The Netherlands}

 \date{received: ; accepted: }
\maketitle

\begin{abstract}
Beginning with a simple model for the growth of structure, I consider
the dissipationless evolution of a MOND-dominated region in an
expanding Universe by means of a spherically symmetric N-body code.
I demonstrate that the final virialized
objects resemble elliptical galaxies with well-defined relationships
between the mass, radius, and velocity dispersion.  These calculations
suggest that, in the context of MOND, massive elliptical galaxies 
may be formed early ($z\ge 10$) as a result of monolithic dissipationless
collapse. Then I reconsider
the classic argument that a galaxy of stars
results from cooling and fragmentation of a gas cloud on a time
scale shorter than that of dynamical collapse.
Qualitatively, the results are similar to that of the traditional
picture; moreover, the existence, in MOND, of a density-temperature 
relation for
virialized, near isothermal objects as well as a mass-temperature
relation implies that there is a definite limit to the mass
of a gas cloud where this condition can be met-- an upper limit
corresponding to that of presently observed massive galaxies.
\end{abstract}

\section{Introduction}

Here I reconsider the problem of galaxy formation, 
dissipationless and dissipational, in the context 
of modified Newtonian dynamics, or MOND (Milgrom 1983).
With Newtonian gravity, dissipational processes do not
appear to be necessary to explain the basic structural properties
of elliptical galaxies.  For example, van Albada (1982) demonstrated 
that the observed surface brightness 
profile of elliptical galaxies, the $r^{1/4}$
law (de Vaucoleurs 1948), develops naturally in Newtonian 
N-body simulations
of the collapse of bound systems with initially inhomogeneous density
distributions.
On the other hand, dissipation does seem to be necessary to explain
the typical densities of baryonic matter in both disk and
elliptical galaxies (Binney 1977); pure gravitational collapse
in an expanding Universe produces insufficient concentration
of visible matter. Moreover, cooling and fragmentation
are certainly necessary for the formation of stars which are
the principal component of galaxies.  Thus,
the formation of galaxies could, in some sense, 
be viewed as dissipationless if
cooling of a primordial gas cloud and fragmentation down to
the level of stellar mass objects occurs on a time scale
short compared to that of dynamical collapse.  This is a point
first made by Hoyle (1953) to explain the mass scale of
galaxies, and developed further by
a number of authors (Silk 1977, Binney 1977,
Rees \& Ostriker 1977, White \& Rees 1978). The argument involving
these two competing timescales has become an
essential ingredient of the standard cosmology in which the matter
budget of the Universe is dominated by a hypothetical
pressureless, collisionless fluid, cold dark matter, or CDM
(White \& Rees 1978, Blumenthal et al. 1984). 

In the context of the MOND paradigm, galaxy formation must occur without
the assistance of CDM.  Moreover, recent numerical work has indicated
that dissipationless 
merging of galaxies proceeds much more slowly with MOND than
when luminous galaxies are assumed to be surrounded by massive
dark halos (Nipoti, Londrillo \& Ciotti 2007b); 
therefore, mergers would probably play
a less important role in galaxy formation, particularly
the formation of ellipticals, than in the current conventional
picture.  But the primary 
roadblock to the consideration of galaxy formation with MOND
is that there is not yet a generally accepted or
standard cosmological context for structure formation,
even though relativistic
extensions have been described in the literature (e.g.,
Bekenstein 2004, Sanders 2005,
Zlosnik, Ferreira \& Starkman 2006).
Therefore, I begin with the simple assumptions that MOND only
applies to peculiar accelerations in an expanding Universe and
that the acceleration constant, $a_0$, does not evolve with
cosmic time.

Such assumptions would appear to be consistent with the current
relativistic versions of MOND, where the non-Newtonian force is
mediated by a long-range scalar field with non-standard Lagrangian:
there is no MOND in the absence of density variations. 
Moreover, these assumptions underly the heuristic models of
structure formation in a MONDian universe considered previously
by myself (Sanders 2001) and by Nusser (2002).  In particular, Nusser noted
that small fluctuations grow rapidly ($\propto (z+1)^{-2}$), and that 
this leads
to a unacceptably clumpy Universe at the present epoch-- essentially
independent
of the initial amplitude of fluctuations.  I postpone consideration
of this problem to a later discussion on taming the growth of 
large-scale structure formation
and assume here that galaxy scale fluctuations in a baryonic Universe
grow by the application of the MOND formula to peculiar accelerations.
If so, then it is easily demonstrated that galaxy mass objects re-collapse
early ($10 < z < 25$) and that spherically symmetric
dissipationless collapse leads to final virialized objects having
roughly the properties of observed elliptical galaxies.  In particular,
both the magnitude and form of the observed surface density distribution 
--the de Vaucoleurs law-- and the velocity dispersion-baryonic mass 
relation-- the Faber-Jackson law (Faber \& Jackson 1976)-- are reproduced for
bound objects with masses ranging up to $10^{13}$ $M_\odot$.

All of this, however, begs the question of why galaxies, as self-gravitating
ensembles of stars, have an upper mass limit to their baryonic
content of about $10^{12}$ $M_\odot$.
More massive self-gravitating objects exist, but they are
bound groups or clusters of galaxies rather than single entities
consisting of stars.
To address this question we must again consider the processes of
radiative cooling and fragmentation on a time scale short compared
to that of gravitational collapse.
In this respect, there is a great advantage afforded by
modified dynamics:  the presence of an additional physical constant
with units of acceleration ($a_0\approx 10^{-8}$ cm/s$^2$) in the
structure equation strongly constrains the properties of any
self-gravitating object as we see in this and previous
dissipationless collapse calculations (Nipoti, Londrillo \& Ciotti
2007a).  Such an
object with a velocity dispersion of a few hundred km/s will have a
mass of $10^{11}$ M$_\odot$ as well as a characteristic size and
density.  In other words, there is less arbitrariness in specifying
initial properties of a pre-galactic cloud.

In the context of MOND the virial mass, the dynamical timescale,
and the characteristic density of a self-gravitating isothermal cloud
depend only upon velocity dispersion.  This redefines the curve,
in the temperature-density plane, describing the condition that dynamical
and cooling timescales are equal.  As in the standard
model, this curve
neatly separates those objects where dissipation has led to the
formation of stars (galaxies) from those where it has not (groups and 
clusters).
Moreover, the presence of a density-temperature relation for self-gravitating
isothermal clouds leads to definite upper limit to the mass of objects 
in which
the cooling time is less than the dynamical time-- again a characteristic
mass of about $10^{12}$ M$_\odot$.  Therefore, we see that many of
the successes of the standard model for galaxy formation carry over to MOND--
in particular the existence of an upper limit to systems consisting
of stars. Moreover, the location of near isothermal systems in the 
temperature-density plane-- systems ranging from globular
clusters to clusters of galaxies--
plane becomes understandable in terms of MOND.

In the following section I review the heuristic model for the initial
growth of galaxy scale fluctuations in an expanding Universe.
In section 3 I numerically follow the dissipationless
evolution of galaxy-scale spherical regions which collapse out of the 
Hubble flow. Even though dissipationless, these calculations demonstrate 
the scaling relations
which primordial gas clouds may obey. 
In section 4 I apply these relations to reconsider the old argument
on collapse and fragmentation in the context
of MOND, and in the final section I compare MOND with the
standard model and summarise.

\section{The growth of small fluctuations in a MONDian Universe} 

MOND in its original form (Milgrom 1983) is described by the
relation between the true gravitational acceleration, $g$, and
the Newtonian gravitational acceleration, $g_N$--
$$g\mu(g/a_0)=g_N\eqno(1)$$
where $a_0$ is the fundamental acceleration parameter ($\approx
10^{-8}$ cm/s$^2$) and $\mu$ is the function which interpolates
between the Newtonian regime ($g>a_0$, $\mu = 1$) and the MOND
regime ($g<a_0$, $\mu = g/a_0$).  MOND, as a modification of
Newtonian gravity, may also be expressed by the modified
field equation of Bekenstein \& Milgrom (1984):
$$\nabla\cdot[\mu(\nabla\phi/a_0)\nabla\phi] = 4\pi G\rho.\eqno(2)$$
This formulation is conservative but difficult to solve for
arbitrary mass distributions;  in the case of spherical symmetry it
reduces to the simple expression above (eq.\ 1) where $g_N$ is 
determined by the usual Poisson equation.
 
Early efforts to describe a MONDian cosmology in the absence of 
a relativistic theory applied the simple MOND
equation (eq.\ 1) to an expanding spherical region 
analogously to the Newtonian
derivation of the Friedmann equation (Felten 1984).
There it was immediately realized that the evolution depended
upon the physical size of the region (it was not possible to
define a dimensionless scale factor) and that
any expanding region would eventually
re-collapse regardless of its initial density and expansion velocity.  However,
at early epochs, the region within which the Hubble deceleration is less than
$a_0$, is much smaller than the horizon scale; therefore, if $a_0$
is independent of cosmic time, it may
be possible that the Universe as a whole is Friedmannian while increasingly
larger sub-regions become MONDian and re-collapse (Sanders 1998). Although
this would lead naturally to a scenario of hierarchal structure
formation, it is not possible to identify the centres of re-collapse;
i.e., primordial density fluctuations play no role.

A more plausible scenario would be one in which the MOND formula 
is applied to the peculiar accelerations developing from density fluctuations
rather than to the Hubble expansion as a whole.  This was the central
idea behind heuristic 
models for structure formation (Sanders 2001, Nusser 2002) and
for the early collapse of low-mass gas clouds \cite{sk05}.
This assumption is consistent with, but not a necessary consequence of,
current relativistic extensions of MOND (for 
example, TeVeS, Bekenstein 2004).
Its validity depends upon the form of the free function of the theory
(effectively the MOND interpolating function) in the cosmological regime
where the cosmic time derivative of the scalar field dominates
the scalar field invariant.  But since the free function does not
follow from any more fundamental considerations, assumptions
about its form, at present,
are no less ad hoc than applying the MOND prescription 
directly to peculiar accelerations.

With the additional ansatz that the MOND acceleration parameter
$a_0$ is does not vary with cosmological time (also not necessarily
true in relativistic extensions such as the biscalar variant of
TeVeS, Sanders 2005), the equation for the growth of small 
fluctuations (($\delta = \delta\rho/\rho$) becomes:
$$\ddot{\delta} + 2{{\dot{x}}\over {x}}\dot\delta + {{\ddot{x}}\over x}\delta
= {{3g_1}\over{x\lambda_c}}\eqno(3)$$
Here $x$ is the dimensionless scale factor in terms of the present
scale factor ($x(t_0) = 1$), time is in units of the Hubble time
($1/H_0$), $\lambda_c$ is the co-moving scale of the perturbation,
and $g_1$ is the peculiar gravitational acceleration related
to the Newtonian peculiar acceleration 
$$g_p = {{\Omega_m}\over {3x^2}}\lambda_c\delta \eqno(4)$$
by the MOND formula (eq.\ 1).  Here I have assumed that fluctuations
grow primarily during the period when non-relativistic matter
with density parameter $\Omega_m$ dominates the mass-energy
budget of the Universe.  Below I will assume, consistent with MOND,
that this matter is essentially baryonic.

In the high acceleration limit
($g_1>a_0$) we recover the usual linear Newtonian expression for the
growth of small fluctuations in an expanding medium ($g_1 = g_p$).  However,
when $g_1<a_0$, then the MOND limit applies:
$$g_1 =\Bigl[{{f_m{\Omega_m\lambda_c r_H}\over{3x^2}}\delta}\Bigr]^{1\over 2}
\eqno(5)$$
where $r_H$ is the Hubble radius, $c/H_0$, and the MOND acceleration
parameter is written as $a_0=f_mcH_0$ with $f_m\approx 1/7$.
In this case, the equation for the growth of small fluctuations 
becomes non-linear ($g_1\propto \delta^{1/2}$) and dependent 
upon the co-moving or mass scale.  

In the MOND limit a simple power law solution to
this equation exists in the matter dominated regime; i.e.,
$\delta=At^\alpha$ with $\alpha=4/3$.  In other words, unlike the
Newtonian case where $\delta\propto x$, in the MOND regime
$\delta \propto x^2$ (Nusser 2002).  
In terms of the scale factor, the solution is
$$\delta = {4\over{27}}f_m{r_H\over \lambda_c}{{x^2}\over
{\Omega_m}}.\eqno(6)$$
In fact, as may occur in
non-linear problems, this solution is an attractor on a space of
solutions;  MOND drives structure growth toward $x^2$ quite independently
of the initial conditions. 

\begin{figure}
\resizebox{\hsize}{!}{\includegraphics{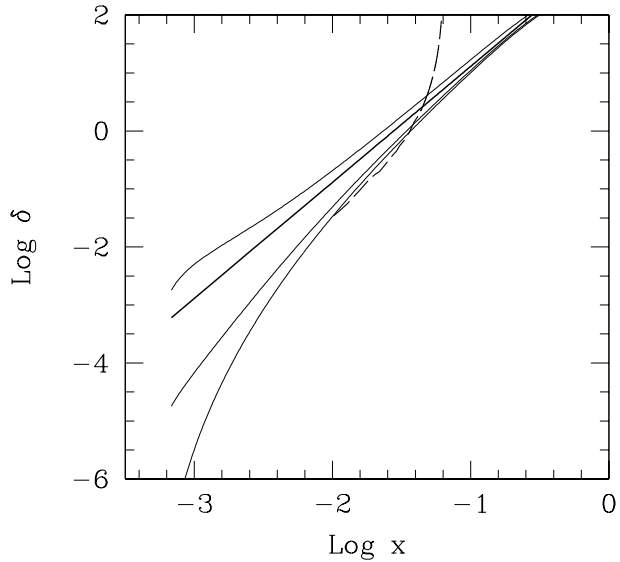}}
\caption[]{The amplitude of a galaxy scale density fluctuation
($\lambda_c=1.56$ Mpc) as a function of scale factor on a log-log
plot.  The heavy solid line is eq.\ 6, the power-law solution of
eq.\ 3, and the
light solid curves are the numerical solutions of the growth equation
(eq.\ 3) in the low density ($\Omega_m=0.04$), vacuum energy dominated
cosmology for initial values at decoupling: $\delta_0 = 1.8\times 10^{-7},\,
1.8\times 10^{-5},\, 1.8\times 10^{-3}$.  The dashed curve is the
evolution of over-density of the expanding sphere determined
by the numerical N-body program; i.e., when $\delta>1$ eq.\ 6 is no
longer valid.} 
\label{}
\end{figure}

This is shown in Fig.\ 1, where $\delta$ is plotted as a function
of scale factor.  The cosmological
background here is of the usual Friedmann form for
a Universe without dark matter
($\Omega_m = 0.04$) but with the standard radiation content and
zero spatial curvature, i.e., $\Omega_\lambda = 0.96$. Note
that here I do not consider the possible presence of 2 eV neutrinos; this
would change details-- on the scale of galaxies, adding an 
additional contribution to Hubble expansion-- 
but not the essence of the argument.  
In eq.\ 1 I have taken the form of $\mu$ to be that
advocated by Zhao \& Famaey (2006), i.e., $\mu(x)=x/(1+x)$.
The heavy 
solid curve is the pure MOND power law solution (eq.\ 6) for
a co-moving scale of 1.56 Mpc corresponding to a mass of $10^{11}$
$M_\odot$, and 
the lighter solid curves are numerical solutions of eq.\ 3 with various
initial values for $\delta$ at the epoch of matter decoupling
($z\approx 1400$).  We see that the numerical solutions rapidly
converge to the power law solution.

It is also evident that the fluctuations grow to unity on a
at relatively early epoch:
$$x_1=2.6\Bigl[{{\Omega_m}\over{f_m}}{\lambda_c\over {r_H}}\Bigr]^{1\over 2}
\eqno(7)$$ corresponding to a redshift of
$z_1 = 13.8(10^{11} M_\odot/M)^{1\over 6}{\Omega_m}^{-{1\over 3}}\approx 
35$ for the galaxy scale fluctuation.  In this scenario we would expect
massive galaxies to form early ($z>10$).  

Given that the power spectrum of fluctuations
is related to the amplitude as $P(k)k^3 \propto \delta^2$ this gives
a final power spectrum of $P(k)\propto k^{-1}$ as noted by
Nusser (2002).  However, also as noted by Nusser, eq.\ 6
would certainly imply that the amplitude of fluctuations on large
scale is far too large to be consistent with observations.
Cosmology can intervene to tame the growth of larger scale structure;
essentially, vacuum energy (or curvature) moderates
the growth when it begins to dominate the cosmic expansion, but this
will be considered in a later paper.  For now, we assume that eq.\ 6 applies
to galaxy scale fluctuations.  Of course, the equation is only 
valid when $\delta<1$; thus below
we consider the dissipationless evolution of collapsing spheres with MOND;
i.e., only (modified) gravity affects the evolution.  The underlying
implicit assumption is that the baryonic fluid, presumably a gas
cloud at the beginning of re-collapse, rapidly cools, fragments
and forms stars during the initial collapse.  The condition
for this to occur will be considered in Section 4.

\section{The expansion and re-collapse of MOND dominated fluctuations}

Fully three-dimensional dissipationless collapse calculations
have recently been carried out by Nipoti et al. (2007a).
Starting with an inflated Plummer sphere (zero initial 
kinetic energy) they follow the collapse and formation of
virialized objects with varying central concentration,
ranging from deep MOND to Newtonian.
Here I use a spherically symmetric N-body code of the 
form originally developed by H\'enon (1964) in order to
follow the evolution of an initially expanding 
over-dense spherical region beyond $\delta = 1$.
Although the calculations of Nipoti et al. would provide 
more realistic models of actual galaxies, the goal here
is to consider the general properties of objects which might actually
condense out of the Hubble flow.

The radial
motion of spherical shell $i$ at radius $r_i$ is determined
by numerically solving the equation of motion:
$${{d^2r_i}\over{dt^2}} = -g_i + {{{j_i}^2}\over{{r_i}^3}}\eqno(8)$$
where the gravitational force at the $i^{th}$ shell, $g_i$,
is given by two parts:  $g_i = g_H + g_1$ where $g_H$ is the usual
Hubble deceleration and $g_1$, as above, is the peculiar deceleration
resulting from the over-density and given by the MOND formula (eq..\ 1),
again with the Zhao-Famaey interpolating function.  Whenever
the peculiar acceleration exceeds the Hubble acceleration by a factor
of two, I assume that the region is decoupled from the Hubble
flow and then follow its evolution as an isolated object.
The final term in eq.\ 8 is the centrifugal acceleration with 
$j_i$ being the specific angular momentum of the $i^{th}$ shell; 
i.e., $j_i = v_tr_i$ where $v_t$ is the tangential velocity of particles
comprising that shell.  Note that there is no systematic rotation;  $v_t$
of particles in a shell is assumed to be distributed uniformly in 
all tangential directions.
Again, because of the perfect spherical symmetry
the solution of eq.\ 1 is equivalent to solving the Bekenstein-Milgrom
modified Poisson equation.

The sphere is initially homogeneous and 
partaking in the uniform Hubble
expansion perturbed by its peculiar velocity.
An initial over-density of $\delta\rho/\rho=0.03$
is simulated by taking the sphere to be initially slightly smaller
($\delta r/r = 0.01$) than it would be in a uniform Universe.
The initial conditions for the expanding sphere then are taken from
the numerical solution for $\delta$ described above (as in Fig.\ 1)
in the case where the fluctuation amplitude at decoupling
is $\delta_0=1.8\times 10^{-7}$.
For the sphere with mass $10^{11}$ $M_\odot$,
when $\delta=0.03$ we find $t_i= 0.00278$ in units of the Hubble time,
corresponding to $x_i= 0.00968$ which yields an initial radius of
15.1 kpc and an expansion velocity of 254 km/s (this is
all for the adopted cosmological model). 
The expansion is taken to be linear ($=rH$) out to
the maximum radius.  The initial tangential velocity on all shells
is assumed to be 20-30 km/s in order to prevent a collapse which is too
violent (the results are qualitatively insensitive to the actual value of 
$v_t$).  With MOND, the  over-dense 
sphere will inevitably re-collapse after
reaching a maximum radius about 3 times larger than the initial
radius.  The re-collapse occurs after a time interval
comparable to the age of the Universe at the epoch when the sphere
decouples from the Hubble flow. 
Of course, the assumptions of uniformity and spherical
symmetry are  enormous simplifications; since smaller regions re-collapse
earlier the region forming a galaxy would
consist of numerous collapsing or virialized sub-components.

\begin{figure}
\resizebox{\hsize}{!}{\includegraphics{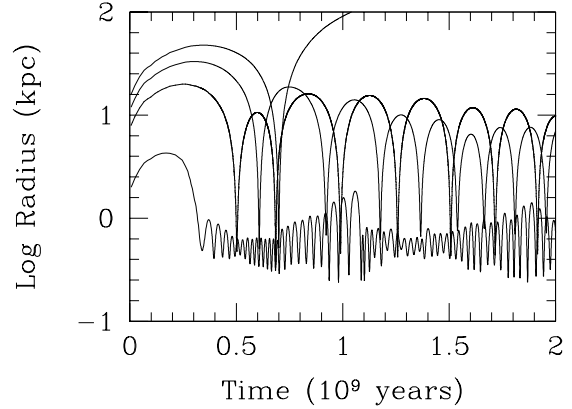}}
\caption[]{The log radius of four characteristic shells in a $10^{11}\ms$ 
sphere, as a function
of time.  Initially the entire sphere is uniformly expanding and,
in a Newtonian context, would be unbound; i.e., the sphere would never
re-collapse.  But with modified dynamics, all shells will eventually 
re-collapse.  After the initial re-collapse,
the shells oscillate at different frequencies, phase mix, and come
into equilibrium.}
\label{}
\end{figure}

This idealised evolution is illustrated in Fig.\ 2 for the $10^{11} 
M_\odot$ sphere simulated by 800 spherical shells.  This shows the 
logarithm of the radius of four different
shells as a function of time;  the largest shell is the outermost shell.
It is evident that the shells oscillate with
different frequencies and inter-penetrate.  Due to the phase mixing, the
sphere will eventually come into an equilibrium state.
The over-density is calculated by comparing the average density within
in the outermost shell to that of the homogeneous Universe at 
that epoch.  This is shown by the dashed curve in Fig.\ 1 where we see
that evolution of $\delta$ follows that of the numerical solution
of eq.\ 3 until $\delta\approx 1$ and then diverges.

The approach to final equilibrium can be seen by considering 
the virial ratio ($2T/V$) as a function of time.  In modified dynamics
the gravitational potential is, properly speaking, not defined;  an infinite
energy is required to move any shell to infinity.  None-the-less
it is possible to define a virial relation which, in equilibrium, takes
the form $2T - V = 0$ where $T$ is the total kinetic energy as usual and,
in spherical symmetry, $$V = -4\pi\int{\rho(r)g(r)r^3 dr}\eqno(9)$$
(Romatka 1992, Gerhard \& Spergel 1994).
Here, $g(r)$ is the modified gravitational acceleration 
given by eq.\ 1.  For the system of
spherical shells this becomes $V = \sum r_im_i g(r_i)$ over all shells.
As in Newtonian dynamics the condition $2T/V = 1$ corresponds to 
equilibrium, but, unlike Newtonian dynamics, if $2T/V > 2$ the system
remains gravitationally bound and will re-collapse.

\begin{figure}
\resizebox{\hsize}{!}{\includegraphics{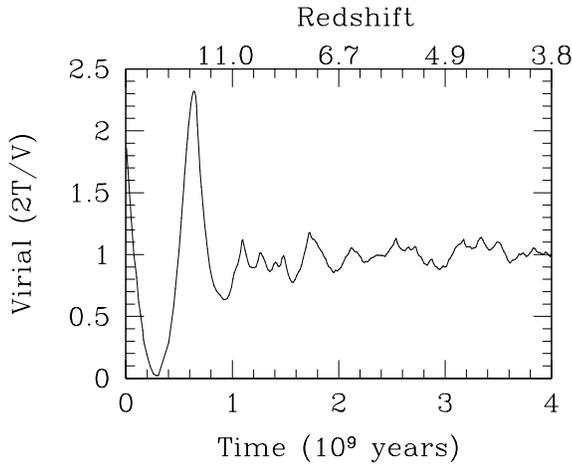}}
\caption{The virial ratio (2T/V) as a function of time for the 
$10^{11}$ M$_\odot$ spherical protogalaxy of Fig.\ 7 (2T/V = 1
in equilibrium).  This illustrates the rapid approach to virial
equilibrium after entering the MOND regime and re-collapsing.
The spherical galaxy is in place as a virialized system 
$2\times 10^9$ years or by a redshift greater
than 6 in the low density universe.}
\label{}
\end{figure}

The virial ratio as a function of time for the $10^{11} M_\odot$ sphere
is shown in Fig.\ 3.  Initially, there is too much kinetic energy
in expansion as would be the case in a low $\Omega$ universe;  the sphere
would never re-collapse in the context of Newtonian dynamics.  
But with modified dynamics re-collapse and mixing
do occur and after roughly three or four dynamical time-scales ($\approx 10^9$
years), the virial ratio approaches one.  Fairly significant 
($\approx 10\%$) oscillations continue for another three to four
dynamical timescale-- somewhat longer than in Newtonian
collapse calculations as found by Ciotti et al. (2007).

The cosmic time and redshift of the initial collapse
of objects of different mass
is shown in Fig.\ 4 as a function of mass.  The initial
conditions and final state of these objects is given in
Table 1.   The initial collapse is that point where
the virial ratio $2T/V$ reaches its first maximum corresponding
to excessive kinetic energy primarily in tangential motion.
We see that initial re-collapse for galaxy mass objects 
($10^{10}-10^{12}$ $M_\odot$) occurs almost coevally, when the cosmic 
age is
between 50 and 100 million years, or at redshifts between 10
and 20.  By redshifts of 8 to 10, these galaxy scale masses would
be in place as virialized objects.

\begin{figure}
\resizebox{\hsize}{!}{\includegraphics{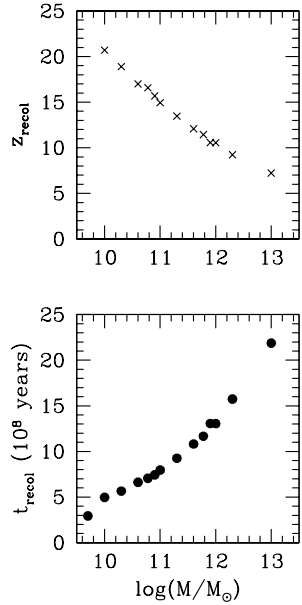}}
\caption[]{The redshift and cosmic age corresponding to initial
re-collapse (defined as maximum virial ratio $2T/V$) for masses
ranging from $10^9$ to $10^{13}$ $M_\odot$.  Galaxy scale objects 
all re-collapse between redshifts of 10 and 20.}
\label{}
\end{figure}

For determining the density and velocity distributions
the effective number of shells is increased
to 8800 by considering the position and velocities of shells at 
different epochs.
The density distribution resembles that of a Jaffe model
(Jaffe 1983) having a power law of exponent near 2 in the inner region 
steepening to more than 3 in the outer regions (we recall that this is
also similar to the density distribution in the MOND isothermal sphere:
Milgrom 1984, Sanders 2000).
The radial velocity dispersion gradually declines as a function of radius, 
as in the high order polytropic spheres considered by
Sanders (2000). This is roughly consistent
with decline observed in elliptical galaxies, at least, out to 
the effective radius. 

\begin{table*}
 \begin{minipage}{10.5cm}
  \centering
   \caption{Initial and final properties of spherical collapse 
        calculations.  (1): mass of sphere ($10^{11}$ M$_\odot$;
    (2): initial cosmic time (current Hubble time); (3) initial
   radius in kpc;  (4) initial velocity at outer radius (km/s);
   (5): redshift at re-collapse; (6) final los velocity dispersion
  (km/s); (7): final effective radius in kpc.} 

\begin{tabular}{@{}lcccccc}

    \hline 
     Mass ($10^{11}$ M$_\odot$) & $t_i$ ($10^{-3}$ ${H_0}^{-1})$ & $r_i$ 
      (kpc) & 
    $V_{max}$ (km/s) & $z_{recoll}$ & $\sigma_0$ (km/s) & $r_{eff}$ (kpc) \\ 
    (1) & (2) & (3) & (4) & (5) & (6) & (7) \\ 

    \hline
    0.05  &  1.66  & 5.6 &  100.2 & 29.1 & 72.1 & 1.75 \\
    0.10  &  1.86  & 7.0 &  135.7 & 20.7 & 95.8 & 3.75 \\
    0.20  &  2.10  & 8.9 &  163.9 & 18.9 & 108.0. & 5.25 \\
    0.40  &  2.37  & 11.2 & 198.0 & 17.0 & 139.4 & 7.0 \\
    0.60  &  2.54  & 12.8 & 221.1 & 16.6 & 155.7 & 8.5 \\
    0.80  &  2.68  & 14.1 & 239.1 & 15.7 & 172.7 & 10.8 \\
    1.0   &  2.78  & 15.2 & 254.0 & 15.0 & 188.3 & 11.5 \\
    2.0   &  3.15  & 19.1 & 306.5 & 13.5 & 231.5 & 16.3 \\
    4.0   &  3.58  & 24.1 & 369.6 & 12.1 & 278.0 & 23.0 \\
    6.0   &  3.86  & 27.5 & 412.3 & 11.4 & 320.7 & 26.8 \\
    8.0   &  4.06  & 30.3 & 445.6 & 10.6 & 346.4 & 30.5 \\
    10.0  &  4.23  & 32.7 & 473.3 & 10.5 & 350.1 & 33.3 \\
    20.0  &  4.82  & 41.2 & 570.4 &  9.2 & 446.3 & 49.5 \\
    \hline
   \end{tabular}
 \end{minipage}
\end{table*}

Because of the large collapse factor the final equilibrium form
is extremely anisotropic with an effective anisotropy radius of
1 kpc-- well within the effective radius.  Such objects would probably be 
highly unstable due to the extreme
anisotropy, but this is suppressed here by the imposed spherical symmetry.
The non-linear development of this instability would lead to
a triaxial figure and a larger anisotropy radius consistent with stability,
as is the case in the three-dimensional collapse calculations of
Nipoti, Londrillo \& Ciotti (2007a).
For this reason, the equilibrium objects resulting from these 
spherically symmetric calculations
are not appropriate in detail as models for elliptical galaxies.  
Nonetheless, it is of interest that the global properties resemble those
of the anisotropic polytropes applied previously (Sanders 2000) 
as models for elliptical galaxies:
specifically, a mild deviation from isothermality similar to that of n=12 to
n=16 polytropes and a velocity distribution which is increasingly 
anisotropic in the outer regions.

\begin{figure}
\resizebox{\hsize}{!}{\includegraphics{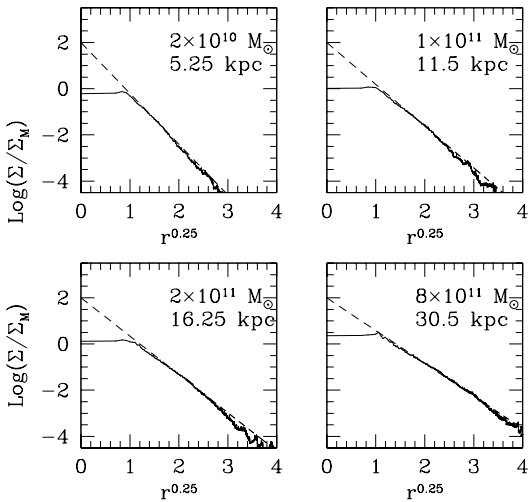}}
\caption{The figures show the surface density distribution for four
different MOND collapse simulations corresponding to the indicated
masses.  In all cases the simulation begins with the sphere in near
Hubble expansion having
an over-density of $\delta = 0.01$ where
the initial conditions are given by the numerical solution of eq.\ 6.
The plots are of log surface brightness vs. $r^{0.25}$ and
the dashed lines indicate the best $r^{0.25}$ law fit with the
effective radius indicated.  The equilibrium figures
are reasonably approximated by the de Vaucouleurs law.  
The flattening in the central regions is artificial and due to
the unrealistically large assumed tangential velocity on
the spherical shells.  The projected central
surface brightness, however, is the same in each case.}
\label{}
\end{figure}

In Fig.\ 5 the surface density profiles are shown for the
equilibrium figures resulting from four different calculations appropriate
to different mass spheres.  We see that these profiles are well fit
by empirically successful $r^{1/4}$ law typically over two orders of
magnitude in surface density.  The flat central surface density
is an artifact of the unrealistically large initial tangential
velocity on shells.  Significantly,
the projected central surface density
is constant, and the effective radius grows, roughly, as the
square root of the mass.  The effective radius is also comparable to
the critical radius for modified dynamics ($r_m=\sqrt{GM/a_0}$);  i.e., 
galaxy mass objects naturally re-collapse to a radius of about 10 kpc
without dissipation. 

It is of some interest that spherically symmetric expansion and
re-collapse in MOND seems to be capable of producing final
virialized objects in which the surface density distribution is
a reasonable approximation to a $r^{1/4}$ law.  With Newtonian
dynamics, it is necessary to begin with a highly
inhomogeneous, or clumpy, initial density distribution in order to
drive the redistribution in phase space sufficient to
achieve such a universal surface density distribution;
uniform spherical collapse does not work (van Albada 1982). 
Of course, it is impossible to repeat the 
calculations described here with Newtonian dynamics because such
spheres would be unbound; there would be no re-collapse.  But beginning
with cold spherically symmetric Newtonian collapse of objects of the
same mass and size scale as described here, I find that the final
range of phase space covered by the particles is considerably more
restricted than for the MOND calculations.  The MOND re-collapse produces
a number of shells which would be unbound in the context of Newtonian
dynamics; this appears to make the critical difference for
the final surface density distribution.

Fig.\ 6 shows the results of the 13 simulations covering a range of masses 
between $5\times 10^9$ and $2\times 10^{12}$ $M_\odot$ on the log effective
radius-log central velocity dispersion plane compared to the 
observations of
J{\o}rgensen et al.\cite{jor99,jfk95a,jfk95b}.  
In several of these simulations a different value
of the initial tangential velocity was assumed ($v_t$ = 20-40 km/s).
Again we see that these global characteristics 
of objects which condense out of the cosmological expansion 
via the MOND prescription are
generally similar to those of actual elliptical galaxies, although
the models are somewhat more inflated than actual elliptical galaxies
(larger effective radius for a given velocity dispersion).  This 
suggests that some global dissipation may be necessary to reproduce
the observed distribution of real objects on this plane.

Fig.\ 7 is the mass-velocity dispersion relation for these 
13 equilibrium objects;
also shown is that of the anisotropic n=12 to n=16 polytropes applied
previously (Sanders 2000) as MOND models for elliptical galaxies which
reproduce the observed fundamental plane. For the collapse models
the velocity dispersion is that along a single line-of-sight
toward the centre weighted by density.  The observed line-of-sight
velocity dispersion within a circular diaphragm, for example of
1.6 kpc diameter \cite{jfk95a}, would be about 25\% lower.

It is evident that the
collapsed spheroids do exhibit a mass-velocity dispersion relation
similar to that observed, although
they are more homologous than actual ellipticals (less scatter on
the Faber-Jackson relation).  This is because the spherical collapse 
model is highly idealised (i.e., no deviation from spherical 
symmetry or internal structure).  These objects would also
lie on a fundamental plane similar to that described
by the anisotropic polytropes (Sanders 2000, eq.\ 15)  
i.e., $M/(10^{11}) {\rm M_\odot} \approx 10^{-5} [\sigma({\rm kms}^{-1}]^{1.76}
[r_e ({\rm kpc})]^{0.98}$ (the scaling is lower due to the
extreme and unrealistic anisotropy). 
But the important conclusion is that objects resembling the 
ellipticals do condense out of the Hubble flow
within the framework of the MOND scenario for structure formation.  
Moreover, the final virialized objects
attain the mean binding energy of actual galaxies without the
necessity of dissipation.  These are not deep MOND objects
(such as dwarf spheroidals or other low-surface brightness
systems) but are more similar to average massive
ellipticals which are Newtonian within an effective 
radius. (Deep MOND isothermal spheres with large constant
density cores are
possible configurations, but not by this formation scenario;
Milgrom 1984, and private communication.)

In summary, objects formed from MOND dissipationless collapse
exhibit well-defined radius-velocity dispersion and
mass-velocity dispersions relationships.
These are of the form
$$r_e = p\sigma^2/a_0\eqno(10)$$
and
$$\sigma^4=qGMa_0 , \eqno(11)$$
where $p\approx q\approx 1$ (shown by the solid lines in Figs. 6 and 7.
These objects differ from pure isotropic,
isothermal spheres where $p=4.36$ and $q=0.0625$
(Milgrom 1984, Sanders 2000).  Significantly, this means that such
objects also have a characteristic density which is also
related only to the velocity dispersion:
$$\rho_e = {1\over{2\pi q p^3}}{{a_0}^2\over{G\sigma^2}}\eqno(12)$$

\begin{figure}
\resizebox{\hsize}{!}{\includegraphics{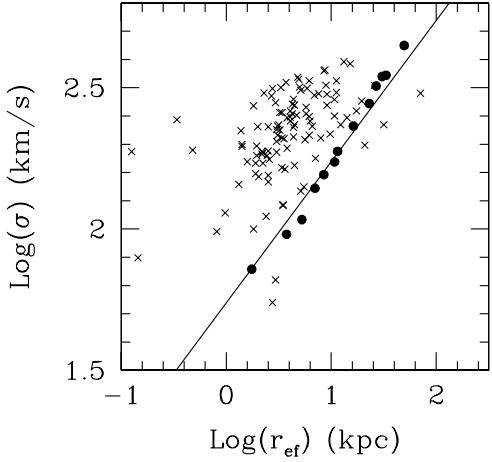}}
\caption{The dark points show the distribution of the equilibrium
collapse models on the log($\sigma_o$)-log($r_e$) plane. 
The cross points are observed 
elliptical galaxies from the samples of J{\o}rgensen
et al.  The solid line is eq.\ 10 with $p=1$.}
\label{}
\end{figure}

\begin{figure}
\resizebox{\hsize}{!}{\includegraphics{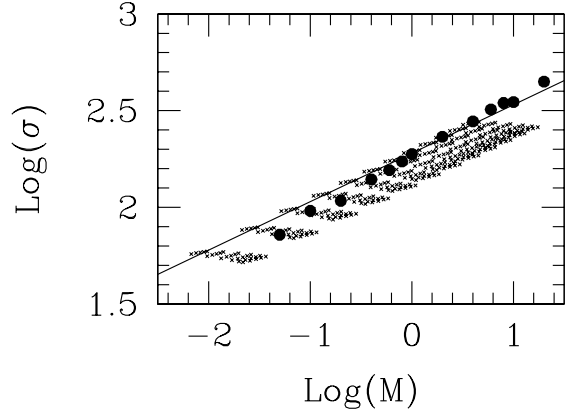}}
\caption{The dark points show the mass-velocity dispersion relation
for the equilibrium collapse models; the fainter crosses are the same
for the anisotropic MOND polytropes applied previously (Sanders 2000) 
as models of elliptical galaxies. The solid line is eq.\ 11 with
$q=1$.}
\label{}
\end{figure}

\section{The MOND condition for fragmentation}

These dissipationless calculations still do not address the question
of why galaxies of stars seem to be restricted to mass of less
than $10^{12}\ms$.  For this we have to consider cooling and
fragmentation processes which lead to star formation.

In the traditional picture, this process is considered in terms
of two competing timescales-- the dynamical time vs. the cooling
time.  Presumably, the elements of the collapsing cloud collide,
heat up, and attain a temperature appropriate to the virial
velocity dispersion. 
The radiative cooling timescale of this hot plasma is
$$t_c = 3kT[\mu m_p \Lambda(T)n]^{-1}\eqno(13) $$
where $T$ is the temperature, $\mu$ is the mean molecular
weight, $m_p$ the mass of the proton, $n$ the density,
and $\Lambda(T)$ is the cooling rate per hydrogen nucleus
in a plasma having specified, presumably primordial, 
abundances.  With Newtonian dynamics,
the dynamical, or collapse, timescale is
$$t_d\approx(G\rho)^{-{1\over 2}}\eqno(14)$$ where $\rho$ is the density.

\begin{figure}
\resizebox{\hsize}{!}{\includegraphics{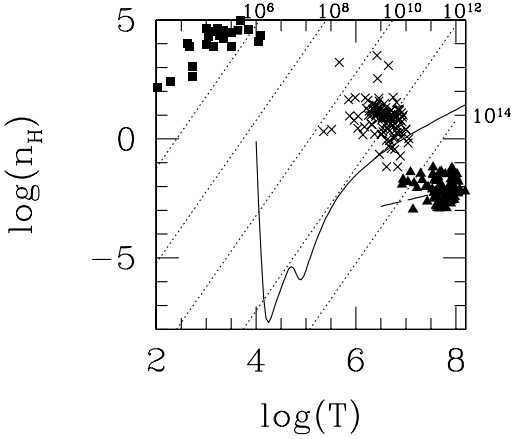}}
\caption[]{The solid curve shows the defines the locus on
the density-temperature plane where radiative cooling time equals
the classical, Newtonian dynamical timescale.  Gas clouds in the
 region above the curve can cool and fragment before the cloud
collapses.  Here one would expect to form self-gravitating objects
consisting of stars.  The parallel dotted lines are the locus
of homogeneous objects of indicated mass in virial equilibrium.  The
squares, crosses and triangles show, respectively, globular clusters,
elliptical galaxies and X-ray emitting clusters of galaxies.
The long dashed line is the locus of cooling time equal to Hubble
time;  cooling flow clusters should lie above this line.}
\end{figure}

Fig.\ 8 is a density-temperature plane for a hypothetical gas cloud
and the solid curve shows the locus where $t_c=t_d$. 
Above the solid curve, one would expect cooling, fragmentation and
star formation to play an important role.  The points on the curve
show three classes of pressure supported systems: globular clusters
(Pryor \& Meylen 1993, Trager et al. 1993), 
elliptical galaxies \cite{jor99,jfk95a,jfk95b} 
and X-ray emitting clusters of
galaxies (White, Jones \& Forman 1997).  
The point made by numerous authors (e.g., Silk 1977) is clear:  
Those objects
consisting primarily of stars, ellipticals,
are above the curve; whereas clusters
of galaxies, consisting primarily of hot gas, are below.

If we consider gas clouds uniformly mixed with a dark matter
halo in virial equilibrium we can also define a relation
between the mean density and the temperature for a given dark matter mass:
$$\rho = {{3f_b}\over{4\pi\alpha^3}}\Bigl({{kT}\over{\mu m_p}}\Bigr)^3
G^{-3}M^{-2}\eqno(15)$$
where $\alpha$ is a number depending upon the density distribution
in the cloud and $f_b$ ($\approx 0.15$) 
is the baryon to dark matter density ratio,.  
This density-temperature relation is also shown in Fig.\ 8
for objects of the indicated total dark mass.  The point is that
for objects with $M>10^{12}\ms$ the virialized cloud will lie
primarily in the region where cooling is slow compared to
the collapse timescale.  In this region we would expect that
the fragmentation to the level of individual stars does not occur before 
subsequent dynamical evolution of the object.

With MOND we have seen that dissipationless objects which condense out of
the Hubble flow exhibit
a well defined radius-velocity dispersion, mass-velocity dispersion
relations (eqs.\ 10 and 11) which imply a density-velocity relation
(eq.\ 12).  However, the initial re-collapsing object, at maximum expansion, 
is certainly
a gas cloud; as the cloud collapses subcomponents will collide
and, most likely, generate a temperature with thermal velocity
comparable to the random velocity of components.  Therefore, a
critical assumption here, as in the standard model, 
is that the gas virial temperature generated
during re-collapse, fragmentation, and star formation is
essentially given by the random velocity of the virialized object.
For our hypothetical gas cloud, therefore, the radius-temperature
and mass-temperature 
relations corresponding to eqs. 10 and 11 are (for $p=q=1$)
$$r_e = 4.5\Bigl({T\over{10^6K}}\Bigr)\, {\rm kpc}\eqno(16)$$
$${M\over{10^{11}\ms}} = 0.14 \Bigl({T\over{10^6K}}\Bigr)^2\eqno(17)$$
where I have taken the mean atomic weight of 0.62 for an ionised
gas with primordial abundances.  The dynamical timescale for the cloud would
therefore be $t_d = p\sigma/a_0$ or
$$t_d=3.8\times 10^7\Bigl({T\over{10^6K}}\Bigr)^{1\over 2}
\,{\rm years}\eqno(18)$$
Expressing the characteristic density of a collapsed object (eq.\ 12)
in terms of the hydrogen density for a fully ionised
plasma with primordial abundances:
$$n_H=0.9\Bigl({T\over{10^6K}}\Bigr)^{-1}\, {\rm cm^{-3}}\eqno(19)$$

\begin{figure}
\resizebox{\hsize}{!}{\includegraphics{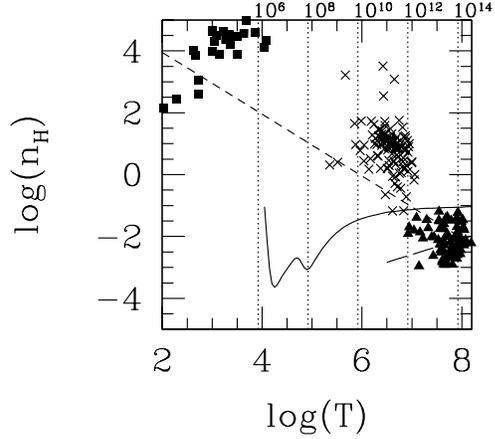}}
\caption[]{The solid curve shows the defines the locus on
the density-temperature plane where radiative cooling time equals
the MOND dynamical timescale.  As above, gas clouds in the
 region above the curve can cool and fragment before the cloud
collapses.  Here one would expect to form self-gravitating objects
consisting of stars.  The vertical dotted lines are the locus
of homogeneous objects of various mass in virial equilibrium
in the context of MOND.  As before, the
squares, crosses and triangles show, respectively, globular clusters,
elliptical galaxies and X-ray emitting clusters of galaxies.  Note that 
the vertical scale differs from that of Fig.\ 8.}
\end{figure}

The condition $t_c=t_d$ where now $t_d$ is defined by eq.\ 18 is
shown by the solid curve in Fig.\ 9.  The three classes of self-gravitating
objects are are indicated here as is Fig.\ 8.  The temperature of
the gas clouds with indicated masses, as given by eq.\ 17, is now 
shown by the vertical dotted lines
(there is no dependence on scale, and hence density,
in the MOND virial relation).  And finally, the characteristic MOND
density (eq.\ 19) for a near-isothermal
cloud in equilibrium is shown by the dashed line.

We see that the solid curve again separates those objects built
primarily from stars (elliptical galaxies) from objects consisting
primarily of gas (clusters).  We also see that the 
MOND density-temperature relation
intersects this curve for an object with a mass
of about $10^{12}$ $\ms$.  This means that objects which
are more massive will have a density and temperature such that
cooling and fragmentation occurs on a timescale longer than
subsequent evolution of the system as a whole; this provides
a natural upper limit for objects built of stars.  It is also evident
that actual virialized systems do cluster about the line of
the characteristic MOND density.  This has been pointed out previously
in other contexts (Sanders \& McGaugh 2002).

If cosmological neutrinos have a mass of 2 eV 
they will comprise the dark component
of clusters required even in the context of MOND (Sanders 2003).  
In this case 
the neutrino fluid will become a dominant component of
clusters at temperatures higher than about $4\times 10^7$ K,
and the curve of $t_d=t_c$ would steepen at higher temperatures
(Sanders 2007).
This does not affect the argument on fragmentation.  If the
dark component of clusters is baryonic (Milgrom 2007), then
the form of the curve remains the same.

At high temperature, the solid curve in Fig.\ 9 
($t_c=t_d$) appears to approach a constant value of the density.
This is because the cooling timescale for a high temperature
plasma is set by free-free emission and is given by
$$t_c ={K {(kT)^{1\over 2}}\over{\mu^2 n_e {m_e}^{1\over 2}
{r_t}^2 {c^2} \alpha}}\eqno(20)$$
where $K$ is a numerical constant, $m_e$ is the mass of
the electron, $r_t$ is the Thompson radius ($e^2/m_e c^2$) and
$\alpha$ is the fine structure constant ($e^2/\hbar c$).  That is to
say, the cooling timescale is proportional to $\sqrt{T}$, as
is the dynamical timescale (eq.\ 18).  Therefore, the temperature
drops out of the condition $t_c=t_d$ in this high temperature regime;
there is a characteristic density where this condition is met of
$n\approx 0.1$ cm$^{-3}$.  Equating this to the
MOND density for a virialized system (eq.\ 19)
then defines a characteristic
temperature or (from eq.\ 17) a mass.  This mass would be the
upper limit to an object where fragmentation to stars has proceeded--
the characteristic galaxy mass.  This turns out to be
$$M_c\approx {\alpha^6}{{\alpha_g}^3}\Bigl({{m_p}\over {m_e}}
\Bigr)^3
  {{\hbar a_0}\over {c^3}} \approx 10^{12} \ms \eqno(21)$$
where $\alpha_g = G{m_p}^2/{\hbar c}$ is the gravitational coupling constant.
Therefore, with MOND, as in the standard paradigm, the galaxy
mass is defined in terms of fundamental constants although, in our
case, $a_0$ enters into the expression.
Given that, coincidentally, $\hbar a_0/c^3 \approx 0.1\alpha_g m_e$, eq.\ 15
becomes $$M_c \approx 0.1\alpha^6{\alpha_g}^{-2} \Bigl({{m_p}\over {m_e}}
\Bigr)^2 m_p\eqno(22)$$
which is similar (but not identical) to the expression derived by Silk (1977).

\section{Conclusions}

It has been appreciated for many years (e.g. Binney 1977) 
that, by the mechanism of gravitational instability in an expanding 
Universe, it is impossible to form the luminous parts of galaxies 
without dissipation. This remains true in the context of
the current cosmological paradigm, $\Lambda$CDM.  Given the
amplitude of fluctuations on galaxy scale at decoupling, bound
objects of $10^{12}$ $M_\odot$ would form at a redshift of about 5
with a characteristic scale of to 20 to 50 kpc (defined as
the radius at which the power law density distribution breaks
from -1 to -3) 
and a density between
$10^{-24}$ and $10^{-25}$ gcm$^{-3}$-- independently of whether or
not they form hierarchically or by monolithic collapse.  
(Navarro, Frenk \& White 1997).  But this would be 
the presumed characteristic density of dark matter.
If the baryonic matter were distributed
in the same way (no dissipation) then its density
would be 10 times lower.  The baryonic 
matter must dissipate, radiate its
kinetic energy and collapse by a further factor of 10 to produce
the observed baryonic density in the inner luminous regions of galaxies.
Moreover, this dissipation must be global; fragmentation to the
level of stars before collapse would not produce galaxies of
the observed surface brightness.

With MOND, given the simple ansatz that the modified gravity only
applies to fluctuations, this is not true.  Then, in a pure
baryonic Universe, galaxy scale fluctuations separate out of
the Hubble flow early ($z>35$) and re-collapse and virialize
by a redshift of 10.  The mean surface density and binding energy
of these collapsed objects is comparable to that of observed galaxies.
The standard surface brightness distribution
for ellipticals, the $r^{1/4}$ law, is reproduced as well as the
Faber-Jackson and Fundamental Plane correlations.  
This is all accomplished without dissipation.
Of course, galaxies are self-gravitating ensembles of stars.
So, to form galaxies in this dissipationless limit, cooling
and fragmentation to the level of stellar mass objects must
occur on a timescale short compared to the collapse time.
Therefore, as in the standard picture, radiative cooling
must set the mass scale of galaxies.

The more robust result here, independent of the ansatz on
structure formation, concerns the properties of protogalactic
clouds.
With MOND, unlike the Newtonian case, an acceleration scale enters
the structure equation.  This means that 
a near isothermal object with a given velocity dispersion,
or temperature, will
have definite mass: if that velocity dispersion is 100-200 km/s
the mass will be on the order of $10^{11}$ $M_\odot$
Moreover, there also is a definite size scale 
associated with an isothermal object of a given mass-- and that is
the radius of the transition from Newtonian dynamics to modified 
dynamics.  The isothermal object will inflate to this radius and thereafter
truncate.  This implies the existence of a characteristic density
which can be identified with a particular mass;  for example,
a mass of $10^{11}\ms$ will have a density of about one particle
per cubic centimetre.  
For lower masses, or temperatures below $10^7$ K, 
the virialized MOND clouds lie 
in the domain where cooling is more rapid than collapse; here we
expect the cloud to fragment and form a galaxy of stars essentially
by dissipationless collapse.  For higher
temperatures this is no longer the case. The cloud may well maintain
its identity as a gaseous object
until merging with a similar cloud as part of a larger collapsing structure.
It is of interest that observed pressure supported, near isothermal objects do,
at present, lie near the characteristic MOND density-temperature relation.

In summary, with MOND galaxy scale masses are likely to re-collapse and
virialize early ($z>10$).  Spherically symmetric N-body 
calculations indicate that the objects which condense out of the
Hubble flow, resemble actual elliptical galaxies.  This suggests
that elliptical galaxies may form by monolithic collapse
without global dissipation.  The condition that cooling takes place rapidly
compared to collapse, as in the standard scenario, places
an upper limit of about $10^{12} \ms$ on those bound objects which
can consists primarily of stars.

I am grateful to Fran{\c c}oise Combes, Stacy McGaugh, and Moti Milgrom
for very helpful comments on this paper.  I also thank the referee,
Pasquale Londrillo, for a number of very usful comments and criticisms
which greatly improved the content and presentation of this paper.

\end{document}